\documentstyle[11pt,aaspp4,epsf,flushrt]{article}

\def\psim{\lower.5ex\hbox{$\; \buildrel \propto \over\sim \;$}}
\def\gtrsim{\lower.5ex\hbox{$\; \buildrel > \over\sim \;$}}
\def\lesssim{\lower.5ex\hbox{$\; \buildrel < \over\sim \;$}}

\def\g2{\gamma_2}

\begin{document}

\title{Beaming, Baryon-Loading, and the Synchrotron Self-Compton 
Component in Gamma-Ray Burst Blast Waves Energized by External Shocks}

\author{Charles D. Dermer\altaffilmark{1},  
        James Chiang\altaffilmark{2,1}, 
       \& Kurt E. Mitman\altaffilmark{3,1}}

\altaffiltext{1}{E. O. Hulburt Center for Space Research, Code 7653, 
Naval Research Laboratory, Washington, DC 20375-5352}
\altaffiltext{2}{JILA, University of Colorado, Campus Box 440, 
                 Boulder, CO 80309-0440}
\altaffiltext{3}{Thomas Jefferson High School for Science and Technology, 
6560 Braddock Road, Alexandria, VA  22312}

\begin{abstract} 

We present detailed calculations of nonthermal synchrotron and
synchrotron self-Compton (SSC) spectra radiated by blast waves that
are energized by interactions with a uniform surrounding medium.
Radio, optical, X-ray and gamma-ray light curves and spectral indices
are calculated for a standard parameter set that yields hard GRB
spectra during the prompt emission phase. Because no lateral spreading
of the blast-wave is assumed, the calculated temporal breaks represent
the sharpest breaks possible from collimated outflows in a uniform
surrounding medium. Absence of SSC hardenings in observed GRB X-ray
afterglows indicates magnetic field generation toward equipartition as
the blast wave evolves. EGRET detections of 100 MeV--GeV photons
observed promptly and 90 minutes after GRB 940217 are attributed to
nonthermal synchrotron radiation and SSC emission from a decelerating
blast wave, respectively.  The SSC process will produce prompt TeV
emission that could be observed from GRBs with redshifts $z \lesssim
0.1$, provided $\gamma$-$\gamma$ opacity in the source is
small. Measurements of the time dependence of the 100 MeV-GeV spectral
indices with the planned {\it GLAST} mission will chart the evolution
of the SSC component and test the external shock scenario. Transient
optical and X-ray emissions from misaligned GRBs are generally much
weaker than on-axis emissions produced by dirty and clean fireballs
that would themselves not trigger a GRB detector; thus detection of
long wavelength transients not associated with GRBs will not
unambiguously demonstrate GRB beaming.

\end{abstract}

\section{Introduction}

The discovery of X-ray afterglows and optical and radio counterparts
to gamma-ray bursts (GRBs) has enabled redshift measurements of GRB
sources and hosts, thereby confirming the hypothesis that GRBs are
cosmologically distant and therefore very powerful (e.g., Costa et
al.\ \markcite{cea97}1997; van Paradijs et al.\ \markcite{vea97}1997;
Frail et al.\ \markcite{fea98}1998).  The prompt gamma-ray emissions
reach, as in the case of GRB 990123, directional energy releases
$\partial E/\partial \Omega$ as large as $\sim 3\times 10^{53}$ ergs
sr$^{-1}$ (Kulkarni et al.\ \markcite{kea99}1999).  The blast-wave
model successfully accounts for the temporal power-law decays observed
in many X-ray and optical afterglows which were predicted several
years prior to their discovery (Paczy\'nski \& Rhoads
\markcite{pr93}1993; Katz \markcite{katz94}1994; M\'esz\'aros \& Rees
\markcite{mr97}1997) .  In this model, a relativistic blast wave is
energized as it passes through and captures material from an external
medium. The power-law decay results mainly from this energizing
process and the accompanying blast-wave deceleration (see, e.g.,
Vietri \markcite{v97}1997; Waxman \markcite{w97}1997; Wijers,
M\'esz\'aros, \& Rees
\markcite{wmr97}1997).

The degree of GRB blast-wave collimation is a crucial unknown. Breaks
in the temporal indices of the afterglow emissions as a consequence of
beaming are implied by analytic estimates (Rhoads
\markcite{rhoads97}1997, \markcite{rhoads99}1999; Sari, Piran,
\& Halpern \markcite{sph99}1999; Panaitescu \& M\'esz\'aros
\markcite{pm99}1999; Wei \& Lu \markcite{wl99}1999) and numerical
calculations (Moderski, Sikora, \& Bulik \markcite{msb99}1999).
Temporal breaks have been observed in some GRB optical afterglows,
namely GRB 990123 (Kulkarni et al.\ \markcite{kea99}1999) and GRB
990510 (Harrison et al. \markcite{hea99}1999), and have been used to
argue for beaming.
Beaming is important since
knowledge of the degree of collimation is required to determine GRB
source origins.  Neither compact object coalescence scenarios nor
collapsar/hypernova models invoking neutrino annihilation or
magnetohydrodynamic processes make sufficient fireball energy to
account for the largest measured GRB energies without invoking opening
half-angles $\psi \lesssim 10^\circ$ (e.g., Janka et al.\
\markcite{jea99}1999; Popham, Woosley, \& Fryer \markcite{pwf99}1999).

A second crucial question related to the
origin of GRBs is whether the prompt emission results from collisions
between a succession of shells ejected from the GRB engine (Rees \&
M\'esz\'aros \markcite{rm94}1994; Kobayashi, Piran, \& Sari
\markcite{kps97}1997) or is instead due to interactions of a single
impulsive relativistic blast wave with inhomogeneities in the external
medium (M\'esz\'aros \& Rees \markcite{mr93}1993; Dermer \& Mitman
\markcite{dm99}1999).  If a ring of material is formed in stellar
collapse events, then extended GRB ejection events could result,
although greater total energy releases occur for shorter accretion
episodes (Popham et al.\ \markcite{pwf99}1999). Detailed calculations
of coalescence events indicate that the maximum energy output occurs
over time scales of milliseconds (Ruffert \& Janka
\markcite{rj99}1999).  Thus if it is established that the GRB engine
ejects plasma over an extended period of several seconds to tens of
seconds, a collapsar/hypernova scenario would seem to be favored. In
contrast, an impulsive, highly beamed event in a low-density
($n_0\lesssim 1$ cm$^{-3}$) surrounding environment would point to
compact object merger events as the origin of GRBs, because a massive
star progenitor to a collapsar event is probably accompanied by strong
stellar winds and a high-density ($n_0\gg 100$ cm$^{-3}$) surrounding
medium.

In this paper, we present calculations of prompt GRB emissions and
afterglows involving a single impulsive ejection event in the
framework of the external shock model.  Although both colliding shell
(Daigne \& Mochkovitz \markcite{dm98}1998) and external shock
(Panaitescu \& M\'esz\'aros \markcite{pm98}1998; Dermer, B\"ottcher,
\& Chiang \markcite{dbc99}1999) models can reproduce the generic
spectral behavior of GRB pulses and profiles, only the external shock
model has been shown (B\"ottcher \& Dermer \markcite{bd99}1999) to
quantitatively fit the $\gtrsim 1$ s $t_{50}$ duration distribution
and the distribution of the peaks $E_{\rm pk}$ of the $\nu F_\nu$
spectra of GRBs measured with BATSE (Mallozzi et al.\
\markcite{mea95}1995).  Fits to these distributions require a wide
range of $\partial E/\partial \Omega$ values and initial Lorentz
factors $\Gamma_0$. This is not in conflict with relativistic beaming
scenarios and the finding that the values of $E_{\rm pk}$ are
preferentially measured within the triggering range of BATSE, because
BATSE is most likely to trigger on emissions from GRB blast waves with
$E_{\rm pk}$ within its 
observing energy range
(Dermer et al.\ \markcite{dbc99}1999).  Fireballs with a wide range of
baryon-loading parameters are thereby implied. Hence the external
shock model predicts new classes of dirty and clean fireballs that
have not been detected due to design limitations of space-based X-ray
and gamma-ray telescopes (Dermer, Chiang, \& B\"ottcher
\markcite{dcb99}1999). As shown here, the dirty fireballs can produce
transients that could be mistaken for the delayed emissions from
off-axis GRBs.

Prompt and afterglow behaviors from beamed GRBs are studied by
choosing parameters derived from the external shock model that yield
hard spectra in the prompt gamma-ray emitting phase. In Section 2, we
describe our numerical treatment and choice for a standard parameter
set with $\Gamma_0 = 300$. We present detailed calculations of the
afterglow light curves from radio through TeV energies in Section 3
for a jet with $\psi = 10^\circ$, and show calculations of the
spectral and temporal index variations due to beaming.  Predictions
for gamma-ray temporal and spectral variation in the prompt and early
afterglow phases due to the evolution of the SSC component in an
external shock model are presented in Section 4. The external shock
model predicts spectral aging of the SSC component in the early
afterglow, leading to a soft-to-hard evolution in the GeV spectra in
the early afterglow phase.  Light curves at a range of observing
angles for blast waves with $\Gamma_0 = 100$ and 1000 are presented in
Section 5, where properties of long-wavelength transients due to dirty
fireballs and misaligned GRB outflows are compared and contrasted.  We
summarize in Section 6.

\section{GRB Blast-Wave Calculations}

The calculations are based on the code described in the paper by
Chiang \& Dermer (1999), and the model employs the standard blast-wave
physics assumptions (see, e.g., Wijers, M\'esz\'aros, \& Rees
\markcite{wmr97}1997; Panaitescu \& M\'esz\'aros \markcite{pm98}1998;
Sari, Piran, \& Narayan \markcite{spn98}1998 and references therein).
The blast wave is modeled by a relativistically expanding surface
that subtends a constant solid angle fraction $f = \delta \Omega/4\pi$
of the full sky.  The blast wave has initial bulk Lorentz factor
$\Gamma_0$ and total directional fireball energy $\partial E/\partial
\Omega = 10^{54}E_{54}$ ergs/(4$\pi$ sr). We choose $E_{54} = 1$ for
the calculations in this paper.  Although a more complicated geometry
that accounts for lateral expansion of the shell (see Rhoads
\markcite{rhoads99}1999) can be easily implemented, the constant solid
angle assumption produces the sharpest breaks in the light curves
(Moderski et al.\ \markcite{msb99}1999). Sharper temporal breaks could
be produced if $f$ decreases with distance $x$ from the explosion
site, but we consider this prospect unlikely since the pressure of the
external medium would not exceed the jet pressure until the blast wave
decelerates to nonrelativistic speeds. The surrounding external medium
is assumed to be uniform with density $n_0$ = 100 cm$^{-3}$. This
density is intermediate to values ranging from $10^{-3} \lesssim n_0
\lesssim 10^6$ cm$^{-3}$ that are considered in most GRB models.

As the blast wave sweeps up material from the external medium, it is
assumed that a constant fraction $\epsilon_e$ of the swept-up kinetic
energy of the particles is converted to the internal energy of
nonthermal electrons, and that the electrons are injected with a
power-law distribution with index $p$ between minimum and maximum
electron Lorentz factors $\gamma_{\rm min}$ and $\gamma_{\rm max}$,
respectively.  The value of $\gamma_{\rm min}$ is obtained by
numerically solving the expression
\begin{equation}
({p-1\over p-2})\; {\gamma_{\rm min}^{2-p} - \gamma_{\rm
max}^{2-p}\over \gamma_{\rm min}^{1-p} - \gamma_{\rm max}^{1-p}} = 1
+\epsilon_e {m_p\over m_e} (\Gamma - 1)\; ,
\label{gmin}
\end{equation}
which follows from kinetic-energy and number conservation for the
swept-up particles, assuming prompt acceleration and no particle
escape. In order to maximize radiative efficiencies, we assume that
the injected energy is equally shared between the protons and
electrons and let $\epsilon_e = 0.5$. The value of $\gamma_{\rm max} =
\epsilon_{\rm max}(3e/\sigma_{\rm T}B)^{1/2} = 4.6\times
10^7\epsilon_{\rm max}B^{-1/2}({\rm G})$ of the injected electrons is
given with respect to the 
limit on $\gamma_{\rm max}$ that
is obtained by balancing the synchrotron loss time scale with the time
scale for an electron to execute a Larmor orbit (Guilbert, Fabian, \&
Rees \markcite{gfr83}1983; de Jager \& Baring \markcite{djb97}1997),
where $B$ is the magnetic field intensity.  Eq.\ (1) has no solution
if $\epsilon_e$ is too large and $\gamma_{\rm max}$ is too small,
because this prescription may require that more energy than
m$_e$c$^2(\gamma_{\rm max}-1)$ must be given to each electron. We
choose $\epsilon_{\rm max} = 1.0$ in our calculations.

The nonthermal electrons lose energy by synchrotron, SSC, and
adiabatic energy-loss processes, as described by Chiang \& Dermer
(\markcite{cd99}1999).  The angle-averaged synchrotron emissivity
function of Crusius \& Schlickeiser (\markcite{cs86}1986) is used to
compute the synchrotron emissivity, and Jones' (\markcite{jone68}1968)
expression for the Compton emissivity spectrum in the head-on
approximation is used to calculate the SSC component (see also
Blumenthal \& Gould \markcite{bg70}1970).  The value of $B$ is
determined according to the standard equipartition prescription
\begin{equation}
{B^2\over 8\pi} = \epsilon_B \lambda n_0 m_pc^2 (\Gamma^2 - \Gamma)\;,
\label{Bfield}
\end{equation}
where we let the compression ratio $\lambda = 4$ in our
calculations. Blast-wave evolution is self-consistently followed from
momentum conservation by calculating the change in internal
energy due to the added energy of swept-up particles, taking into
account energy losses of the nonthermal electrons. Although electrons
lose energy through adiabatic processes that could be rechanneled into
the kinetic energy of the outflowing blast-wave fluid, we do not
follow this flow of energy. The synchrotron self-absorption
coefficient is calculated according to the standard radiation formulas
as described by Dermer, B\"ottcher, \& Chiang (1999, in preparation),
and radiation spectra with angle-dependent effects of self-absorption
are calculated and summed to produce the observed spectra and light
curves measured at observing time $t$ and photon energy $E$.

The absorption coefficient $\kappa_{\gamma\gamma}(E^\prime)$ to
photon-photon pair production attenuation at comoving frame photon
energy $E^\prime$ is calculated from the formulas of Gould \&
Schr\'eder (\markcite{gs65}1965) and Brown, Mikaelian, \& Gould
(\markcite{bmk73}1973). The emergent spectra are reduced by the factor
$[1-\exp (-\tau_{\gamma\gamma})]/\tau_{\gamma\gamma}$, where
$\tau_{\gamma\gamma}(E^\prime) = \kappa_{\gamma\gamma}(E^\prime)\Delta
x/\cos\xi$, $\Delta x$ is the comoving frame shell width, and $\xi$ is
the angle between the normal to the element of radiating surface and
the observer. We let $\Delta x = x/\Gamma$ (Blandford \& McKee
\markcite{bm76}1976; Panaitescu \& M\'esz\'aros
\markcite{pm98}1998). The reinjection of the pairs and the subsequent
cascade are not followed in this calculation, but the net reduction of
radiated power by the inclusion of this process is found to be a small
($\lesssim 10$\%) fraction of the total radiant power for the
parameters studied here. The received spectra are calculated by
summing over all elements of the radiating surface that contribute to
emission observed at time $t$.


The values we choose for our standard set of parameters are motivated
by observations of GRB emission during the prompt phase.  For example,
as shown by Chiang \& Dermer (\markcite{cd99}1999), this means that
the magnetic field equipartition parameter should be $\epsilon_B
\lesssim 10^{-4}$ during the prompt emission phase in order to avoid
forming cooling distributions which have photon fluxes $\Phi(E)
\propto E^{-3/2}$ below the peak photon energy, $E_{\rm pk}$. 
 Such soft spectra are not commonly observed in GRBs (see Preece
et al.\ \markcite{pea98}1998; Cohen et al.\ \markcite{cohen97}1997).
We note, however, that in order to fit the spectral and temporal
behavior of the {\em afterglow} emission, we require the electron
injection power-law index to be $p \simeq 2.5$ (see, e.g., Wijers \&
Galama \markcite{wg99}1999).  In accordance with the standard blast
wave model, we also assume that $\epsilon_B$ and the other
microphysical parameters $p$, $\epsilon_e$, and $\epsilon_{\rm max}$
do not change with time.  This is obviously an oversimplification, and
could strongly affect afterglow evolution and the importance of the
SSC component in the afterglow phase. In fact our results suggest that
$\epsilon_B$ increases toward equipartition as the blast wave evolves
with time, as explained below.

Figure~\ref{standard_case} shows results for our standard parameter
set with $\Gamma_0 = 300$, $E_{54} = 1$, $n_0 = 100$ cm$^{-3}$, $p =
2.5$, $\epsilon_e = 0.5$, $\epsilon_{\rm max} = 1.0$, and $\epsilon_B
= 10^{-4}$.  Here we show results for an uncollimated blast wave. The
deceleration radius and time scale for these parameters are $x_{\rm
dec} = (3E_0/4\pi n_0 m_pc^2 \Gamma_0^2)^{1/3} = 2.6\times 10^{16}$ cm
and $t_{\rm dec} = x_{\rm dec}/ 2 \Gamma_0^2 c = 9.6$ s, respectively.
The heavy solid curve in Fig.~\ref{standard_case}a shows the
dependence of $\Gamma$ on distance $x$ from the explosion center. This
evolutionary behavior was obtained by an iterative procedure to ensure
self-consistency, and has clearly converged. Because we assume that
$\epsilon_B$ is independent of time and is assigned such a small
value, the GRB spectral properties are in the weak cooling regime
throughout its evolution (Sari et al.\
\markcite{spn98}1998). Even so, the blast-wave does not follow an
adiabatic evolutionary behavior with $\Gamma(x) \propto x^{-3/2}$, but
rather a $\Gamma(x) \propto x^{-1.9}$ behavior in the asymptotic
regime $1 \ll x/x_{\rm dec} \ll \Gamma_0^{1/2}$. Although the lowest
energy nonthermal electrons do not efficiently cool, much of the
energy carried by the higher energy nonthermal electrons is
efficiently radiated away, causing the blast-wave dynamics to depart
considerably from adiabatic behavior.

Fig.~\ref{standard_case}b shows the temporal evolution of the comoving
electron energy spectra $N(\gamma)$ multiplied by $\gamma^2$, where
$\gamma$ is the electron Lorentz factor in the comoving blast wave
frame. Fig.~\ref{standard_case}c shows the calculated $\nu L_\nu$
photon spectra. The various curves in Figs.~\ref{standard_case}b and
\ref{standard_case}c range from 0.01 s to $10^8$ s in factors of 10,
with the later curves showing progressively lower energy breaks. The
lower-energy synchrotron and higher-energy SSC components in the $\nu
L_\nu$ spectra of Fig.~\ref{standard_case}c are evident, and the SSC
component dominates the energy losses of the electrons in much of the
afterglow phase for this set of parameters. SSC processes can be
important when $\epsilon_e/\lambda\epsilon_B \gtrsim 1$ (see Sari,
Narayan, \& Piran
\markcite{spn96}1996; Moderski et al.\ \markcite{msb99}1999 for a more
precise criterion). For the parameters in Fig.~\ref{standard_case},
$\epsilon_e/\lambda\epsilon_B \cong 10^3$, and the SSC component
dominates electron cooling, particularly in the afterglow phase when
the Klein-Nishina effects are less important for electron Compton
scattering. At very late times $t \gg 10^7$\,s, however, the
synchrotron component again dominates when most of the higher energy
electrons have cooled to give a very soft electron spectrum.

Light curves at radio, optical, X-ray, soft gamma-ray, GeV, and TeV
photon energies are shown in Fig.~\ref{standard_case}d in an $L_\nu$
representation. A number of interesting effects are apparent here.
The emission at soft gamma-ray energies rises and decays on the
deceleration time scale due to the energization of the blast wave and
the subsequent deceleration.  The peaking of the TeV light curve due
to SSC radiation mirrors the behavior of the synchrotron peak in the
100 keV light curve.  The optical and X-ray light curves show delayed
emission compared to the soft gamma-ray light curve.  This is due to
the time required for the $\nu F_\nu$ peak of the synchrotron
emission, arising from the lowest-energy cooled electrons in the
electron distribution function, to pass into the various observing
ranges as the blast wave decelerates. The delayed peaking of the lower
energy synchrotron emission represents a prediction of the external
shock model (Dermer et al.\ \markcite{dbc99}1999). As the blast-wave
continues to decelerate, the self-Compton radiation produces
time-dependent flattenings of the light curves at successively later
times for the progressively lower GeV, 100 keV, X-ray, and optical
frequencies. The softenings of the light curves at $t \gtrsim 5\times
10^7$ s occurs when the blast wave reaches nonrelativistic speeds.

The X-ray and optical emissions display temporal behaviors $\propto
t^{-1.75}$ at $t \sim 10^3$--$10^4$ s and $\propto t^{-1.4}$, at $t
\sim 10^5$--$10^7$ s.  A hardening of the temporal behavior of the
X-ray emission occurs at $t \gtrsim 3\times 10^4$ s due to the onset
of the SSC component, and begins to decay more rapidly at $\gg 10^6$
s, when the peak of the SSC component has passed through the X-ray
regime.  Typical observed temporal indices at X-ray and optical
frequencies of GRBs are somewhat flatter than these values; this would
result if a smaller value of $\epsilon_e$ were used in the simulation,
yielding a blast wave evolution more nearly adiabatic, or if a harder
electron index $p$ were used. Lack of evidence for a delayed hardening
from the SSC component in X-ray afterglow light curves suggests that
the magnetic field strength increases towards $\epsilon_B\sim 1$ in
the afterglow phase (see Section 4.1).  These calculations show that
excess emission could
be present in optical afterglow light curves at late times due to the
appearance of the SSC component.  Excess emission has been observed
from GRB 970228 (Reichart \markcite{reichart99}1999) and GRB 980326
(Bloom et al. \markcite{bea99}1999) $\sim 10$--40 days after the
bursts and has been attributed to light from a supernova.  This excess
emission is very red and supports the interpretation of a supernova
origin. To the extent that SSC hardenings have not been observed in
X-ray afterglows, it seems less likely that such SSC hardenings would
be detected in the optical light curves, even taking into account that
optical telescopes have much better $\nu F_\nu$ sensitivities than
X-ray telescopes.  Although a full parameter study has not been made,
caution must be taken to assure that the Compton component from the
blast wave does not introduce delayed excess emission in optical light
curves.  We do note, however, that depending on the physical
parameters, excess emission in the late afterglow from SSC processes
may be bluer at the outset than the synchrotron emission immediately
preceding it and would make a transition over time to a redder
spectrum.  This is clearly seen in the X-ray spectral index shown in
Fig.~\ref{indices}a.  This behavior could then serve as a means of
distinguishing it from supernova emission.

\section{Blast-Wave Calculations of Beamed GRBs}

Figures~\ref{lcs_std}a and \ref{lcs_std}b show angle-dependent light
curves at various frequencies for the same parameter set used in
Fig.~\ref{standard_case}, except now we consider a jet with an opening
half-angle of $\psi = 10^\circ$.  For a one-sided jet, this angle
represents 0.76\% of the full sky. Light curves are evaluated at
inclination angles $\theta = 0^\circ, 10^\circ, 12^\circ, 15^\circ,
30^\circ, 60^\circ$, and 90$^\circ$.  A temporal break due to beaming
occurs in the $\theta = 0^\circ$ curves at $t \sim 10^5$ seconds,
independent of energy, though the break is hidden at MeV and X-ray
energies due to the emergence of the SSC component.  The temporal
break occurs when the angle $\theta \sim
\Gamma^{-1}$ of the Doppler beaming cone equals the jet opening angle
$\psi$.  For a blast-wave decelerating according to the approximation
$\Gamma(x) = \Gamma_0$ for $x < x_{\rm dec}$, and $\Gamma(x) =
\Gamma_0 (x/x_{\rm dec})^{-g}$ for $x > x_{\rm dec}$, the time $t_b$
at which the temporal break is observed with respect to the time of
the explosion is given by $t_b \approx (1+z)[- x\cos \psi + \int_0^x
dx^\prime \beta^{-1}(x)]/c$. Here $c\beta(x)=
c[1-\Gamma(x)^{-2}]^{1/2}$ is the speed of the blast wave.  This gives
\begin{equation}
{c t_b\over (1+z)} = x_{\rm dec}\;[(\Gamma_o\psi)^{1/g} (1-\cos\psi) +
{2g + (\Gamma_0\psi)^{2+1/g}\over 2 (2g+1)\Gamma_0^2} ]\; \;.
\label{ttime}
\end{equation}
Eq.\ ({\ref{ttime}) implies that the temporal break occurs at $t_b
\approx 10^5$ s for $g \cong 1.9$, in agreement with the numerical
results.

Figure~\ref{lcs_std}a shows 8.6 GHz radio, V-band, and 3 keV X-ray
light curves calculated at the values of $\theta$ listed above. The
light curves at $\theta < \psi$ are very similar, as can be seen by
comparing the $0^\circ$ and $10^\circ$ curves, but the X-ray and optical
light curves
drop markedly in intensity once the observer is outside the beaming
cone.  The emission remains at a very low level until the Doppler cone
of the beamed emission intersects the observer's line of
sight. Following the arguments given to derive eq.\ ({\ref{ttime}), we
find that an off-axis observer ($\theta >\psi$) starts to detect
emission at a level comparable to an on-axis observer at times given
by eq.\ ({\ref{ttime}), but with $\psi$ replaced by $\theta-\psi$.

Figure~\ref{lcs_std}b shows light curves calculated at 4.8 GHz radio,
MeV, GeV, and TeV photon energies for the same parameters as used in
Fig.~\ref{lcs_std}a.  No intervening absorption of the GeV or TeV
fluxes are shown in this figure; for cosmologically distant GRB
sources the absorption of TeV photons due to interactions with the
cosmic diffuse infrared radiation field would significantly reduce the
flux compared with the intrinsic emission, depending sensitively on
the redshift of the source (see, e.g., Salamon \& Stecker
\markcite{ss98}1998). The effects of intrinsic $\gamma$-$\gamma$
attenuation are important for the TeV light curves, though not for the
GeV and lower energy light curves, and have been taken into account as
previously described.  In order to illustrate the energy dependence of
this attenuation, we plot in Fig.~\ref{tau_gg} the optical depth to
$\gamma$-$\gamma$ attenuation as a function of observed photon energy
along radial lines-of-sight from the explosion center through the
blast-wave shell.  Each of the optical depth curves corresponds to a
spectrum shown in Fig.~\ref{standard_case}c.  As in
Fig.~\ref{standard_case}c, the vertical lines indicate the energies of
the various light curves we plot in Fig.~\ref{standard_case}d.  It is
evident from this figure that this source of optical depth is only
significant for $\gtrsim 100$ GeV--TeV energies (long-dashed line).

One of the most notable features of Fig.~\ref{lcs_std}b is the
appearance of extended GeV emission compared to the times of the peaks
of the MeV and TeV fluxes.  For the parameters used here, the GeV band
is situated between the peaks of the synchrotron and SSC fluxes, and
the GeV emission is formed from synchrotron emission of the highest
energy electrons in the prompt phase, and primarily from the SSC
process in the early afterglow phase (compare
Fig.~\ref{standard_case}c).  The delayed emission in the GeV band
arises from the same effect that causes the predicted energy-dependent
delays in the peaks of the synchrotron emission at photon energies
$E_{\gamma} \ll E_{\rm pk}$, as described analytically by Dermer et
al.\ (\markcite{dcb99}1999).

An effect (see also Dermer \markcite{Dermer99}1999) revealed in the
calculations shown in Figs.~\ref{lcs_std}a and \ref{lcs_std}b is that
the ratios of the on-axis to off-axis fluxes are much larger at higher
photon energies. Hence the $\gamma$-ray and X-ray synchrotron fluxes
for on-axis observers are very bright in comparison with the fluxes
that would be detected for an identical source that is directed away
from the observer.  This ratio is smaller at optical frequencies, and
the fluxes are comparable for on-axis and misaligned sources at radio
frequencies where the emission persists at comparable levels for about
the same length of time.  This suggests that radio surveys to monitor
misaligned GRB sources would not have to scan over the same region
more than once per day to once per week in order to catch transient
misaligned GRBs. In the case of optical emission, this time scale is
$\sim$ once per hour or so.  The appearance of prompt optical emission
in GRB 990123 (Akerlof et al.\ \markcite{aea99}1999) demonstrates,
however, that an additional prompt optical component, due possibly to
emission from the reverse shock (M\'esz\'aros \& Rees
\markcite{mr93a}1993a; Sari \& Piran \markcite{sp99}1999) which has
not been included in this calculation, could produce brighter and 
more rapidly
varying long wavelength transients.

Spectral and temporal indices are plotted in Figs.~\ref{indices}a and
\ref{indices}b, respectively, for the radio, optical, and hard X-ray
light curves from Figs.~\ref{standard_case} and \ref{lcs_std}a. Here
we write the flux density $F_\nu(\nu,t)\propto
\nu^{-\alpha} t^{-\chi}$ at frequency $\nu$.  The heavy curves show
the $\theta=0^\circ$ case with $\psi = 10^\circ$, and the light curves
show the uncollimated case.  The radio, optical, and X-ray spectral
indices are calculated between 4.8 and 8.6 GHz, the V and I bands, and
3 and 100 keV, respectively. At early times $t \ll t_{\rm dec}$, the
indices all approach $\alpha = -1/3$, corresponding to the $F_\nu
\propto \nu ^{1/3}$ behavior of the elementary synchrotron emissivity
spectrum from an electron distribution function with a low-energy
cutoff $\gamma_{\rm min} \gg 1$.  Only at later times after the
comoving electron density increases does the radio emission become
self-absorbed to form a spectrum rising roughly as $F_\nu \propto
\nu^{-2}$. In the early afterglow phase $1 \ll t/t_{\rm dec} \ll
10^3$, the optical and X-ray spectral indices soften but never
distinctly display an index characteristic of an uncooled distribution
which, for $p = 5/2$, is $\alpha_{\rm uncooled} = (p-1)/2 = 0.75$.
For the chosen parameters, cooling is important for all but the lowest
energy injected electrons.

In the afterglow phase $10^3 \ll t/t_{\rm dec} \ll 10^6$, the optical
spectral indices approaches the value expected for a cooled electron
distribution, namely $\alpha_{\rm cooled} = p/2 = 1.25$. The X-ray
index never reaches this asymptote because the X-ray spectrum hardens
as the SSC component starts to become important in this band at $t
\gtrsim 10^4$ s. The optical spectral index reaches the cooling
spectral index asymptote at $t \gtrsim 3\times 10^4$ s. Beaming
effects on the spectral indices begin to be important at $t \gtrsim
10^5$\,s.

The temporal indices shown in Fig.~\ref{indices}b change in concert
with the spectral indices of Fig.~\ref{indices}a.  The radio emission,
with its inverted spectrum, rises monotonically in flux following the
prompt phase, and the temporal index remains negative (i.e.,
increasing flux) until late times $t \approx 10^7$ s for the case with
no collimation. In contrast, collimation softens the radio temporal
index by about 0.5 units (Rhoads
\markcite{rhoads99}1999; Sari, Piran, \& Halpern
\markcite{sph99}1999). The optical R-band temporal index is $\approx
0$ in the early afterglow phase due to competition between Doppler
deceleration and the increasing number of nonthermal electrons
injected through the sweep-up process. The optical temporal index in
the uncollimated case approaches the value $\chi \sim 1.4$ at
$t\gtrsim 3\times 10^4$ s. For the beamed case, the optical light
curve decays much more rapidly when $t \gtrsim 10^5$ s.  This is also
the case for the X-ray decay, though the appearance of the SSC
component complicates the behavior by introducing a hardening and a
slower decay. This also occurs in the optical spectrum, though at
late times $t \gtrsim10^6$ s.

\section{Gamma Ray Light Curves}

Fig.~\ref{fluences} shows the MeV, GeV, and TeV gamma-ray light curves
of Fig.~\ref{lcs_std}b multiplied by time.  Plotted in this way, the
curves indicate the period during which a gamma-ray telescope with
negligible background will detect most of its counts. As in the
previous section, the heavy and light curves refer to the beamed and
uncollimated cases, respectively.  The pronounced softening of the
light curves at $t \gtrsim 10^5$ s due to beaming is apparent.

\subsection{Hardenings in MeV and X-ray Light Curves during the Early 
Afterglow}

The MeV light curves show strong emission that peaks on the
deceleration time scale --- this is, of course, the GRB itself. A
hardening due to the onset of the SSC component occurs at $t\approx
10^3$ s and reaches a secondary maximum at $t \approx 10^5$ s. The
maximum MeV flux from the SSC component is 4--5 orders of magnitude
less than the peak flux measured in the prompt phase of the GRB (see
Figs.~\ref{standard_case} and \ref{lcs_std}a), but the longer duration
available to accumulate signal will partially compensate if the
background is sufficiently low.

Unfortunately, strong and variable backgrounds at hard X-ray and soft
gamma-ray energies over time scales from hours to days make it
unlikely that an SSC hardening in the MeV afterglow could be observed
with BATSE on the {\it Compton Observatory}.  A very strong GRB with a
signal to noise ratio $\gtrsim 10^3$ could make it possible to observe
the SSC hardening with a pointed instrument. The OSSE instrument on
{\it Compton} has a program to slew toward GRBs.  Persistent emission
decaying as a power-law has been detected as late as $\sim 10^3$ s
after a GRB trigger in two cases, namely GRB 970827 and GRB 97110
(Matz et al.\ \markcite{mea99}1999).  No evidence of hardening in the
light curve was observed at later times, but the S/N ratio may not have
been sufficiently great for detection of delayed emission from these
GRBs.

The lack of distinct SSC hardenings has not been observed in X-ray
afterglows of Beppo-Sax, which our calculations indicate would begin
at $\sim (3\times 10^4$--$10^5)\times (1+z)$\,s after the GRB. Even
though the X-ray afterglow flare occurring $\sim 1$--3 days after GRB
970508 shows a correlated spectral hardening and temporal flattening
(Piro et al.\ \markcite {pea98}1998), we do not think that this is due
to the SSC component because the flaring behavior is more abrupt than
expected from our calculations.  The flare is more likely due to
enhanced emission radiated by the blast wave when it encounters a
density enhancement in the circumburster medium.  The simplest
explanation that no distinct SSC component has been observed in X-ray
afterglows is that $\epsilon_B$ increases from the very low value
found during the prompt phase. Consequently the relative importance of
the SSC component would decrease with time.  Depending on the time
scale over which the magnetic field increases, a hardening in the MeV
component might therefore be still more difficult to detect than
indicated by Fig.~\ref{fluences}.

\subsection{GeV Afterglows and the Case of GRB 940217}

In contrast with the MeV light curves, the prospect of detecting
delayed GeV emission from the SSC component is more favorable because
of the much smaller backgrounds at GeV energies and the earlier onset
of the SSC hardening. The photon fluxes are, however, much smaller at
GeV energies than at MeV energies, so a GeV detector will detect many
fewer GRBs than an MeV telescope with comparable effective
area. Specifically, the EGRET instrument on {\it Compton} has detected
7 GRBs above 30 MeV (Catelli, Dingus, \& Schneid
\markcite{cds98}1998). EGRET has about the same effective area as the
Large Area Detectors on BATSE, but views only 1/20th of the full sky
as compared with 40\% for BATSE.  This indicates that EGRET detects
only $\sim 3$\% of the BATSE bursts that fell within its field of
view, consistent with the much smaller photon fluxes at 100 MeV and
GeV energies compared with the 50--300 keV fluxes measured with BATSE.

The GeV light curve in Fig.~\ref{fluences} shows two peaks.  The early
maximum coincident with the MeV peak is the high-energy extension of
the synchrotron component, and the second maximum peaking at $\approx
5000$ s is due primarily to SSC radiation.  The dimmer synchrotron
flux at GeV energies compared to MeV energies, coupled with the longer
interval over which the relatively brighter SSC signal can be
accumulated, make it more probable that delayed GeV emission rather
than MeV emission can be detected from a GRB, provided of course, that
a sufficiently bright GRB should be observed by a GeV detector.

We propose that the high-energy emission observed 90 minutes after GRB
940217 (Hurley et al.\ \markcite{hea94}1994) was in consequence of the
SSC component becoming increasingly dominant at later times as the
deceleration of the blast wave caused the SSC emission to sweep
through the GeV band. This explanation also accounts for the
appearance of the extraordinary 18 GeV photon observed with EGRET 90
minutes after the onset of the GRB, rather than during the prompt
phase.  The GeV band resides in the $\nu F_\nu$ valley between the
synchrotron and SSC components.  During the prompt phase, the flux of
1--10 GeV photons is very low and the spectrum is very soft;
furthermore there is not much time to accumulate signal.  The
emergence of the SSC component in this band as the blast wave
decelerates maintains the 1--10 GeV flux level at a roughly constant
value (see Fig.~\ref{standard_case}c) over an extended period of time
which, with its harder spectrum, can favor the detection of the
highest energy emission during the early afterglow phase rather than
during the prompt phase.

\setcounter{footnote}{0}

The planned {\it Gamma ray Large Area Space Telescope (GLAST)}
mission\footnote{http://glast.gsfc.nasa.gov/SRD} will have larger
effective area and field-of-view than EGRET, so will likely be able to
monitor the evolution of the SSC spectral feature due to blast wave
deceleration. The time dependence of the MeV--GeV and GeV--TeV 
spectral behavior is plotted in the inset to
Fig.~\ref{fluences}. During the prompt phase, the MeV--GeV photon
spectral index from the cooled synchrotron radiation is $\sim 2.25$,
in accordance with measurements of $> 30$ MeV EGRET spectra (Catelli
et al.\
\markcite{cds98}1998). The GeV--TeV index is much harder because it is
primarily sampling the harder SSC component. As the blast wave begins
to decelerate in earnest during the early afterglow phase, the MeV--GeV
index hardens as the SSC radiation sweeps into this waveband.  We
would then expect that the GeV--TeV index approaches the $\alpha = 5/4$
cooling spectrum, but the effects of $\gamma$-$\gamma$ attenuation on
the TeV emission produces an additional softening. The MeV--GeV index
does approach the cooled synchrotron limit at late times until the
blast wave reaches the nonrelativistic regime.  Spectral hardening in
the MeV--GeV band in the early afterglow phase due to the deceleration
of the blast wave as it interacts with a smooth external medium
constitutes generic behavior of the external shock model which can be
tested with {\it GLAST}.

\subsection{TeV Emission from GRBs}

Figure~\ref{fluences} also shows that TeV photons from the SSC process
during the prompt phase of a GRB is roughly coincident in time with
the prompt MeV synchrotron emission. For this set of parameters, the
$\nu F_\nu$ flux is somewhat dimmer at TeV energies than at MeV
energies. Even though this calculation suggests that strong TeV flux
should be emitted from most GRBs, 
it would only be detected from nearby, low redshift bursts for which
the attenuation due to intergalactic infrared emission is small
(provided that the intrinsic $\gamma$-$\gamma$ attenuation in the
source is also small).  According to the calculations of Stecker \& de
Jager (\markcite{sdj98}1998), the optical depth to TeV photons is
$\cong 0.5$-1 at $z = 0.075$, and increases strongly with increasing
redshift. Thus we should expect TeV emission to be detected only
rarely from a GRB.

Searches for TeV emission from GRBs using the water Cherenkov
Milagrito detector have been recently reported (McEnery et al.\
\markcite{mea99}1999).  By looking for excesses in the $t_{90}$ time
intervals of 54 BATSE GRBs, enhanced emission associated with TeV
radiation was reported from one candidate, GRB 970417a.  The chance
probability for detecting such an excess was calculated to be less
than the $1.5\times 10^{-3}$.  Our calculations are consistent with
detection of TeV emission from GRBs that are close enough to avoid
strong attenuation from the diffuse radiation field.

\section{Baryon-Loading Effects in Beamed GRBs}

Up to this point, we have examined multiwavelength light curves for
beamed and uncollimated outflows employing a standard parameter set
that is consistent with typical durations, spectra, and energy fluxes
observed from GRBs during their prompt phase.  As demonstrated in the
studies by Dermer et al.\ (\markcite{dcb99}1999) and B\"ottcher \&
Dermer (\markcite{bd99}1999) where a simplified analytic form for
emission from blast waves was employed, the spectral properties of
GRBs in the prompt phase are rather insensitive to $n_0$ and $E_0$ ---
varying typically as the cube root of these quantities --- but are
very strongly dependent on $\Gamma_0$ and a parameter related to the
magnetic field.  We have already described how the magnetic field
equipartition parameter $\epsilon_B$ must be assigned a value much
smaller than unity to be compatible with spectral properties of GRBs
during the prompt phase; this may, in fact, indicate the difficulties
of strong field generation during the early episodes of the blast
wave.  There is no {\it a priori} reason to suppose, however, that
$\Gamma_0$ must be assigned a value of 300.

In Figs.~\ref{lcs_dirty} and \ref{lcs_clean} we present calculations
for $\Gamma_0 = 100$ and $\Gamma_0 = 1000$, respectively, with all
other parameters, including $\psi$,
 the same as in the previous sections. The main
differences between the light curves with different baryon loading are
due to the strong $\Gamma_0$-dependences of $t_{\rm dec}$ and the
maximum $\nu L_\nu$ peak flux $\Phi_{\rm pk}$, which is measured at
photon energy $E_{\rm pk}$, the peak of the $\nu L_\nu$ spectrum in
the prompt phase. In accordance with analytical results, $t_{\rm dec}
\propto \Gamma_0^{-8/3}$ (Rees \& M\'esz\'aros \markcite{rm92}1992)
and $\Phi_{\rm pk}\propto \Gamma_0^{8/3}$ (Dermer et al.\
\markcite{dcb99}1999). Thus dirty fireballs with small values of
$\Gamma_0$ are extended and reach smaller peak $\nu F_\nu$ fluxes at
lower values of $E_{\rm pk}$, whereas the clean fireballs with large
values of $\Gamma_0$ produce brief, intense episodes of emission
having larger peak $\nu F_\nu$ fluxes at higher values of $E_{\rm
pk}$.

For the dirty ($\Gamma_0 = 100$) and clean ($\Gamma_0 = 1000$)
fireball cases in Figs.~\ref{lcs_dirty}b and \ref{lcs_clean}b, the MeV
flux maxima occur at $t\approx 63$ s and $t \approx 0.25$ s and reach
maximum flux values of $2\times 10^{49}$ ergs s$^{-1}$ and $10^{52}$
ergs s$^{-1}$, respectively. The MeV flux maximum for the standard
case in Fig.~\ref{lcs_std}b in Fig.~\ref{lcs_clean}b occurs at
$t\approx$ 4 s and $1.5\times 10^{51}$ ergs s$^{-1}$. The product of
the peak flux and the time of flux maximum, which corresponds roughly
to the duration of the GRB, is proportional to the number of counts
that would be measured by a GRB detector; these products for the
dirty, standard, and clean cases are $\sim 10^{51}$, $6\times
10^{51}$, and $3\times 10^{51}$ ergs, respectively. Triggering biases
in an MeV detector therefore strongly favor detection of the standard
case over the dirty fireball due to its lower flux and more extended
duration. Such biases also favor the detection of the standard case
over the clean fireball case, though the level of background in
determining relative detectability of clean fireballs is more
important here than for dirty fireballs.  These detailed
calculations support the analysis of (Dermer et al.\
\markcite{dea99}1999) and B\"ottcher \& Dermer (\markcite{bd99}1999)
that the origin of the peaking $\nu F_\nu$ distribution observed with
BATSE is due to selection biases in GRB detectors and the spectral
properties of relativistic blast waves.

By directly overlaying Figs.~\ref{lcs_std}a, \ref{lcs_dirty}a, and
\ref{lcs_clean}a, it is clear that X-ray emissions from dirty
fireballs which do not trigger a GRB detector are generally much
brighter than the X-ray emissions from misaligned GRBs with $\theta
\gg \psi$. Thus we argue that any detection of so-called ``orphan"
afterglows (see Rhoads \markcite{rhoads99}1999) will more likely be
due to dirty fireballs than off-axis GRBs.

Figure~\ref{prompt_lcs} illustrates the preceding points more
clearly by plotting 100 keV light curves for the standard, dirty and
clean fireball cases. This photon energy corresponds to the typical
energy of photons detected by BATSE within its triggering range (for a low
redshift GRB). A GRB
detector will more likely be triggered by a blast wave which places
the peak of its $\nu F_\nu$ flux in the triggering range of the
detector. The strong detector biases against detecting a long duration
dirty fireball, which reaches a relatively very low maximum flux value
at 100 keV compared to the case with $\Gamma_0 = 300$, are clearly
displayed by the inset.

According to this analysis, burst detectors that trigger in a
particular energy range are more sensitive to GRBs with $E_{\rm pk}$
in the energy range of the detector.  We therefore predict that the
planned {\it Swift} mission, with its Burst Alert
Telescope\footnote{http://swift.gsfc.nasa.gov/instruments/bat.html}
that triggers in the 50-150 keV range, would detect a sample of GRBs
with, on average, longer durations and lower $E_{\rm pk}$ fluxes than
those observed with BATSE.

\section{Summary}

Because of its fewer underlying assumptions and success in explaining
GRB phenomenology during the prompt gamma-ray emission phase, we have
restricted calculations in this paper to the external shock model.
Our calculations follow blast waves that are energized and decelerate
by sweeping up material from a uniform surrounding medium, and we
employ parameters consistent with typical durations, spectra, and
$E_{\rm pk}$ values measured from BATSE GRBs.  We have examined the
effects of blast-wave collimation on the light curves. Although we do
not consider lateral expansion of the blast wave, as recently examined
numerically by Moderski et al.\ (\markcite{mea99}1999), our
calculations improve upon this study by including SSC processes and a
self-consistent treatment of blast-wave dynamics.

Our beaming calculations reproduce effects pointed out in the earlier
analytic and numerical studies, namely that afterglows decay more
rapidly for beamed than for uncollimated outflows. When the SSC
component is unimportant, the spectrum also softens as beaming effects
on the temporal decay become important.  Collimated GRB outflows
directed away from an observer will produce extended transients at
X-ray and optical frequencies at a flux level orders of magnitude less
than transients observed along the jet axis. The relative flux levels
of on-axis and misaligned jet sources at radio infrared and radio
frequencies are much smaller than at optical and X-ray frequencies.
Transients from misaligned GRB outflows could be confused with on-axis
clean and dirty fireball bursts, so detection of such long wavelength
transients will not unambiguously demonstrate beaming.

The calculations of the SSC component demonstrate a number of
potentially observable effects:

\begin{enumerate}
\item{Hardenings of the multiwavelength light curves appear as the SSC 
component sweeps through successively lower observing frequencies at
increasingly later times.  For the standard parameter set used here,
TeV emission is due to SSC radiation in the prompt GRB phase, and the
onset of SSC hardenings at GeV, MeV, X-ray, and optical frequencies
occurs at $\sim 200$, $3\times 10^3$, $10^5$, and $3\times 10^6$ s,
respectively. }
\item{Because distinct SSC hardenings have not been observed in GRB X-ray 
afterglows, we argue that the magnetic field in the blast wave
increases toward its equipartition value as the blast wave evolves. A
stronger magnetic field has the effect of reducing the importance of
the SSC component.  If this interpretation is correct, then the SSC
hardenings would not be expected in late time optical afterglows,
strengthening the likelihood that observed hardenings are due either
to an underlying host galaxy or supernova emission.}
\item{We attribute the prompt and delayed 100 MeV--GeV emissions observed 
from GRB 940217 with EGRET (Hurley et al.\ \markcite{hea94}1994)
primarily to nonthermal synchrotron and SSC emission, respectively.
The external shock model predicts a soft-to-hard evolution of the
spectrum at GeV energies in the early afterglow phase that could be
tested with the planned {\it GLAST} experiment.  Taking into account
internal pair production attenuation, we find that TeV SSC emission at
$\nu F_\nu$ levels comparable that of the prompt MeV synchrotron
radiation could be observed from nearby GRBs with $z \lesssim 0.1$,
where attenuation by the cosmic diffuse infrared radiation field is
small.}
\end{enumerate}

Our summary result of this study is that SSC emission produces a broad
spectral feature that can be used for the study of processes in GRB
blast waves. When analyzed with correlated X-ray and soft gamma-ray
observations, observations of the SSC component might make it also
possible to infer magnetic field strengths and Doppler factors, as has
been done in analyses of relativistic outflows in blazars (e.g.,
Catanese et al.\ \markcite{cea97}1997).  Because the SSC radiation is
most intense during the prompt and early afterglow phases, the GeV
range is an especially important regime for GRB studies because the
deceleration of the blast wave --- or the acceleration of a radiating
surface due to collisions between internal shocks --- can be charted by
monitoring spectral and temporal evolution in this band.

\acknowledgments{This work is supported through the NASA Astrophysical 
Theory Program (DPR S-13756G) and the Office of Naval Research.}

\eject

\noindent \bf{Figure~Captions}
 
\setcounter{figure}{0}

\figcaption{Numerical simulation using standard parameter set (see 
text). (a) Evolution of blast-wave Lorentz factor.  Vertical dotted
lines refer to observer times with respect to the time of explosion in
factors of 10, with the left-most line at $t=1$ s. (b) Comoving
electron energy distribution $dN(\gamma;t)/d\gamma$ multiplied by
$\gamma^2$ for observer times ranging from $t = 10^{-2}$ s to $t =
10^8$ s in factors of 10.  (c) Photon spectra in a $E^2 dN/dtdE$ (or
$\nu L_\nu$) representation for observer times ranging from $t =
10^{-2}$ s to $t = 10^8$ s in factors of 10. (d) Light curves in an
$L_\nu$ representation at 8.6 GHz radio (solid curve), optical R band
(dotted), 3 keV X-ray (dashed), 100 keV soft gamma-ray (dash-dotted),
GeV (dashed/triple-dotted), and TeV (long-dashed) photon energies.
\label{standard_case}}

\figcaption{Light curves calculated at various observing energies and 
inclination angles $\theta$ for a GRB with a standard parameter set
and opening half-angle $\psi = 10^\circ$ of the jet. The initial blast
wave Lorentz factor $\Gamma_0 = 300$. Calculations of $\theta =
0^\circ, 10^\circ, 12^\circ, 15^\circ, 30^\circ, 60^\circ$, and
90$^\circ$ are shown, with the brighter peak fluxes reached by curves
progressively closer to the jet axis. (a) Light curves at 8.6 GHz
radio (solid curves), V-band optical (dotted), and 3 keV X-ray
(dashed) are plotted.  (b) Light curves at 4.8 GHz radio (solid
curves), MeV (dot-dashed), GeV (dotted), and TeV (dashed) are plotted.
\label{lcs_std}}

\figcaption{Optical depth to $\gamma$-$\gamma$ attenuation as a function
of observed photon energy.  The curves shown here correspond to the
the various spectra in Fig.~\ref{standard_case}c. Each curve is
labeled at the left by $\log_{10}(t_{\rm obs})$.  The vertical lines
indicate the energies of the light curves shown in
Fig.~\ref{standard_case}d.  For the standard set of parameters,
photon-photon attenuation is clearly only important for the emission
at energies $\gtrsim 1$\,TeV
\label{tau_gg}}

\figcaption{Energy spectral indices and temporal indices for the 
uncollimated (light curves; from Fig.~\ref{standard_case}) and beamed
(heavy curves; from Fig.~\ref{lcs_std}) cases with $\psi = 10^\circ$
and $\theta = 0^\circ$. (a) Energy spectral indices $\alpha$
calculated between 4.8 and 8.6 GHz, the I and V bands, and 3 and 100
keV. (b) Temporal indices $\chi$ calculated at 8.6 GHz, the R band,
and 3 keV.
\label{indices}}

\figcaption{Product of $\nu L_\nu$ flux and observing time $t$ for the 
MeV, GeV, and TeV light curves using the $\theta = 0^\circ$ case in
Fig.~\ref{lcs_std}b.  Inset shows temporal variation of the broadband
MeV--GeV and GeV--TeV energy spectral indices.
\label{fluences}}

\figcaption{Same as Fig.~\ref{lcs_std}, but for an initial blast wave Lorentz 
factor $\Gamma_0 = 100$.
\label{lcs_dirty}}

\figcaption{Same as Fig.~\ref{lcs_std}, but for an initial blast wave Lorentz 
factor $\Gamma_0 = 1000$.
\label{lcs_clean}}

\figcaption{Prompt and early afterglow 
100 keV light curves for blast wave Lorentz factors 
$\Gamma_0 = 1000$ (solid curve), 300 (dashed), and 100 (dotted).
Inset shows light curves on a linear scale.
\label{prompt_lcs}}

\setcounter{figure}{0}

\begin{figure}
\centerline{\epsfxsize=\textwidth\epsfbox{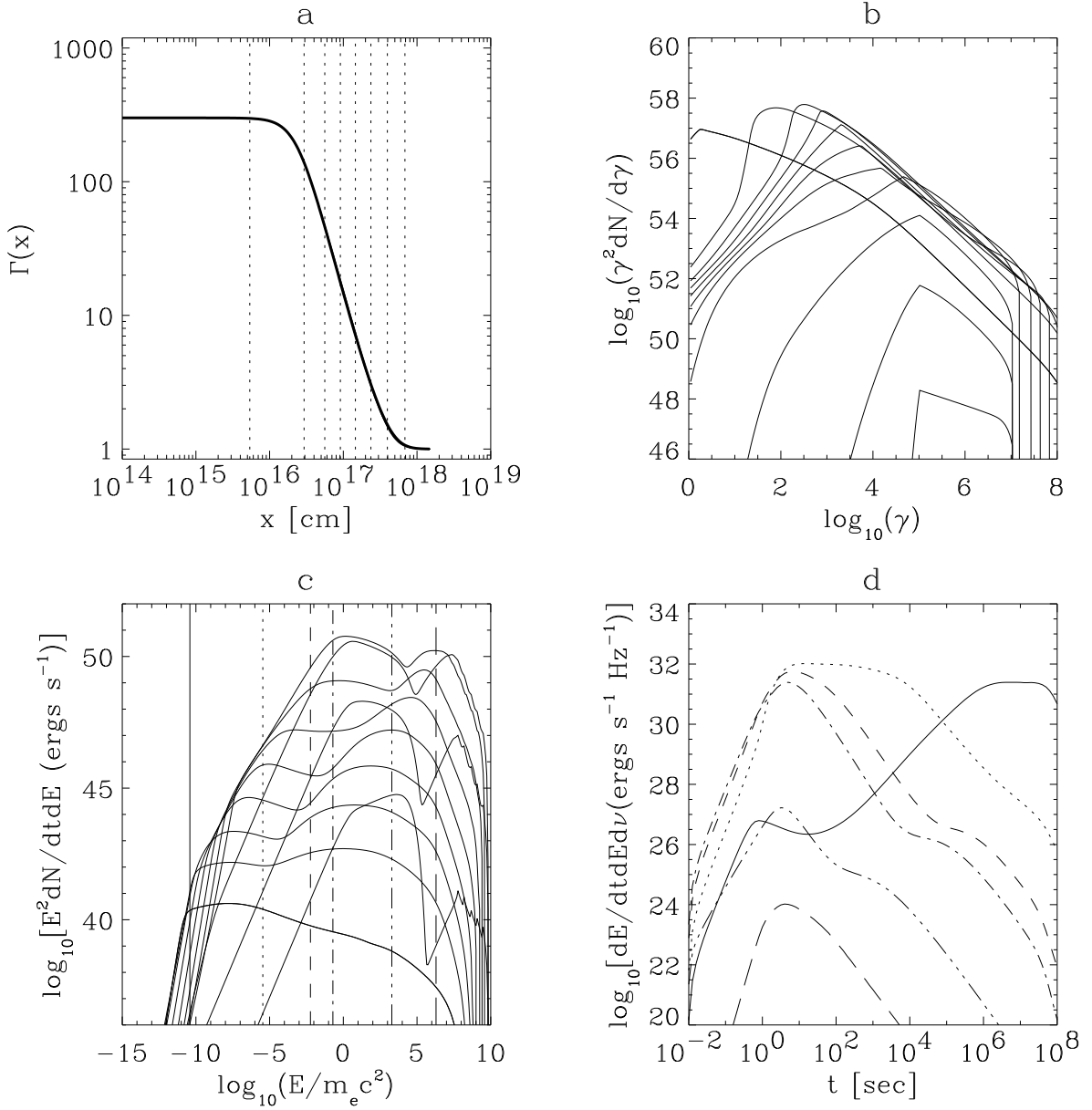}}
\caption[]{Numerical simulation using standard parameter set (see text). 
(a) Evolution of blast-wave Lorentz factor.  Vertical dotted lines
refer to observer times with respect to the time of explosion in
factors of 10, with the left-most line at $t=1$ s. (b) Comoving
electron energy distribution $dN(\gamma;t)/d\gamma$ multiplied by
$\gamma^2$ for observer times ranging from $t = 10^{-2}$ s to $t =
10^8$ s in factors of 10.  (c) Photon spectra in a $E^2 dN/dtdE$ (or
$\nu L_\nu$) representation for observer times ranging from $t =
10^{-2}$ s to $t = 10^8$ s in factors of 10. (d) Light curves in an
$L_\nu$ representation at 8.6 GHz radio (solid curve), optical R band
(dotted), 3 keV X-ray (dashed), 100 keV soft gamma-ray (dash-dotted),
GeV (dashed/triple-dotted), and TeV (long-dashed) photon energies.}
\end{figure}

\begin{figure}
\centerline{\epsfxsize=0.5\textwidth\epsfbox{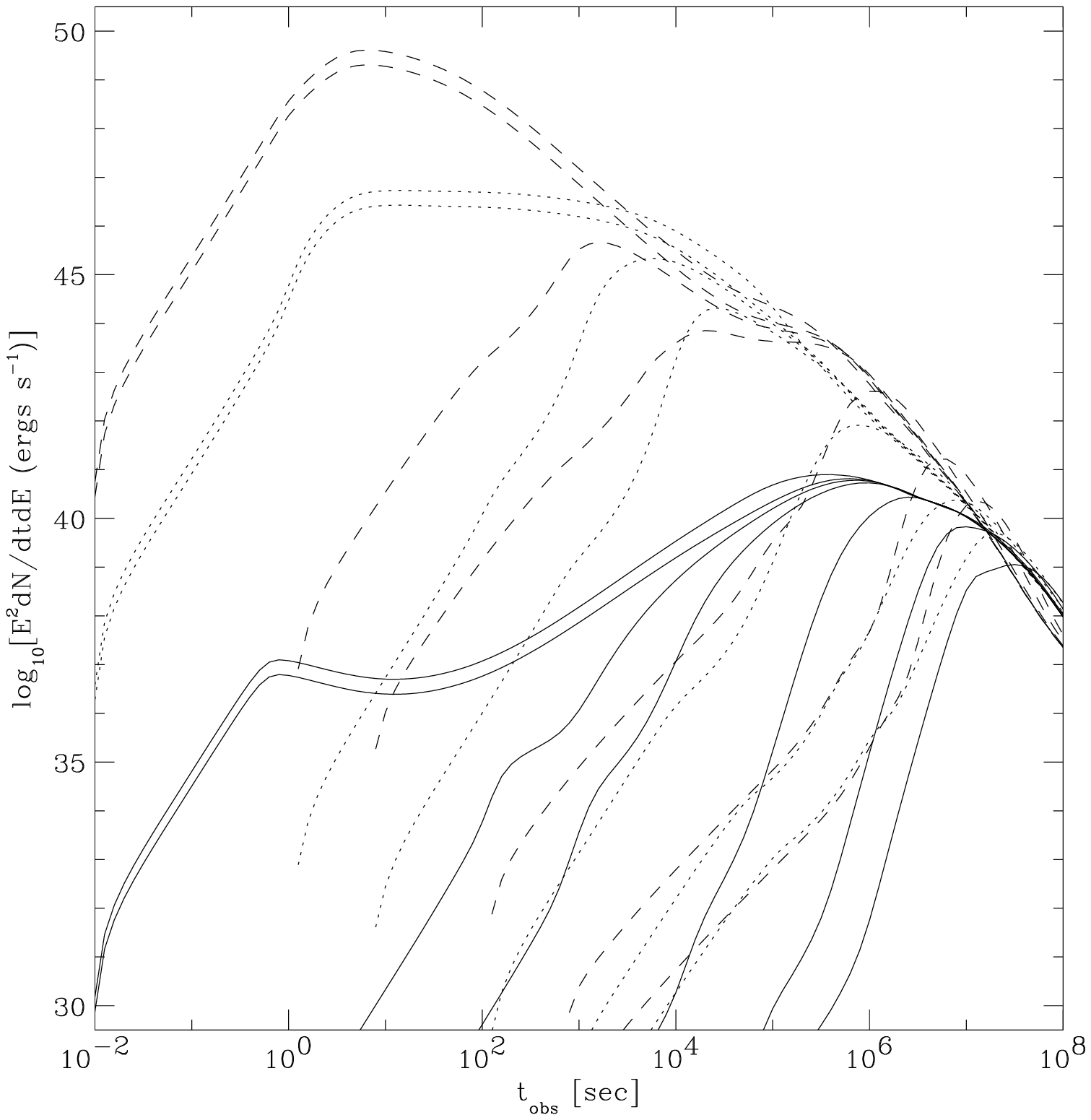}
            \epsfxsize=0.5\textwidth\epsfbox{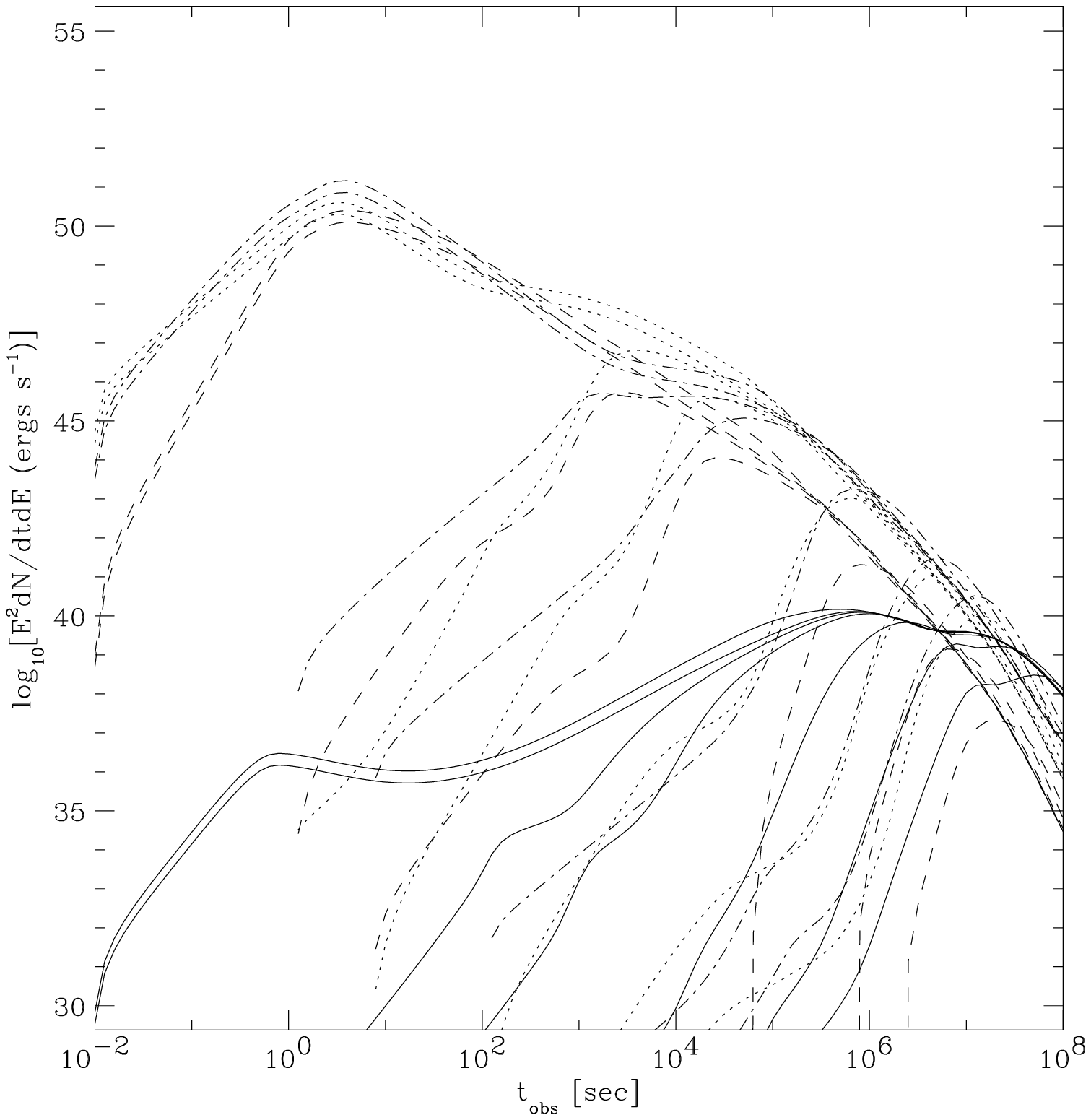}}
\caption[] {Light curves calculated at various observing energies and 
inclination angles $\theta$ for a GRB with a standard parameter set
and opening half-angle $\psi = 10^\circ$ of the jet. The initial blast
wave Lorentz factor $\Gamma_0 = 300$. Calculations of $\theta =
0^\circ, 10^\circ, 12^\circ, 15^\circ, 30^\circ, 60^\circ$, and
90$^\circ$ are shown, with the brighter peak fluxes reached by curves
progressively closer to the jet axis. (a) Light curves at 8.6 GHz
radio (solid curves), V-band optical (dotted), and 3 keV X-ray
(dashed) are plotted.  (b) Light curves at 4.8 GHz radio (solid
curves), MeV (dot-dashed), GeV (dotted), and TeV (dashed) are plotted.}
\end{figure}

\begin{figure}
\centerline{\epsfxsize=0.8\textwidth\epsfbox{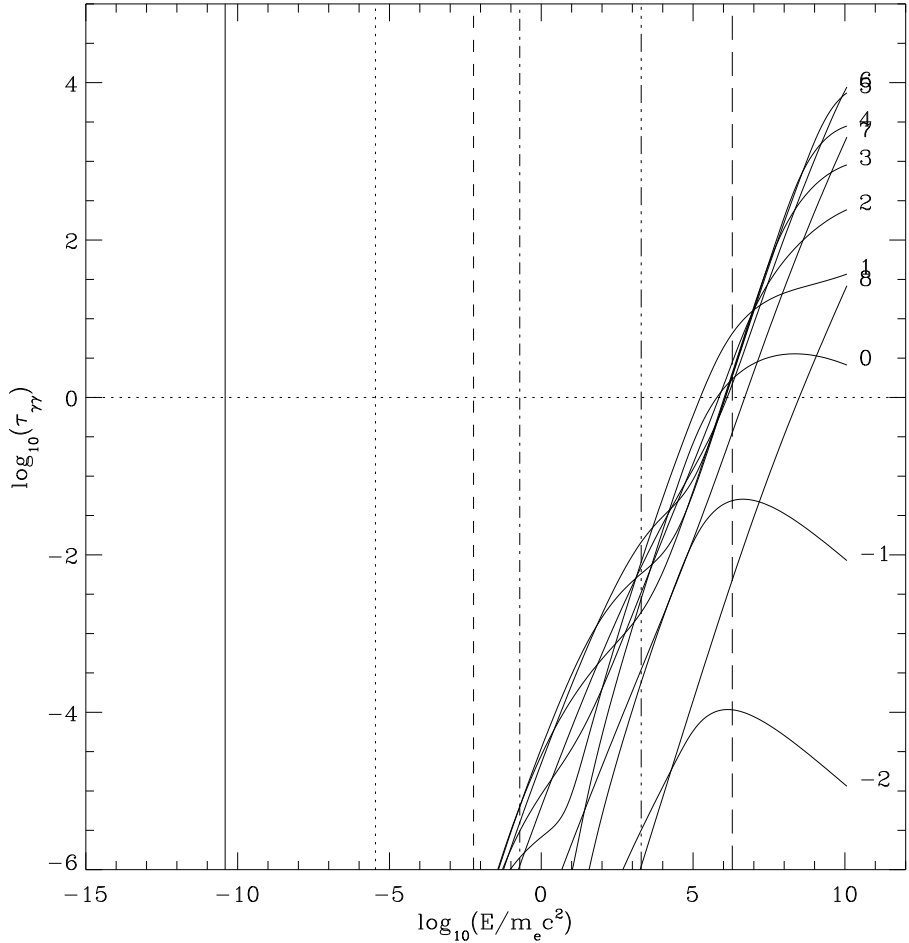}}
\caption[]{Optical depth to $\gamma$-$\gamma$ attenuation as a function
of observed photon energy.  The curves shown here correspond to the
the various spectra in Fig.~\ref{standard_case}c.  Each curve is
labeled at the left by $\log_{10}(t_{\rm obs})$.  The vertical lines
indicate the energies of the light curves shown in
Fig.~\ref{standard_case}d.  For the standard set of parameters,
photon-photon attenuation is clearly only important for the emission
at energies $\gtrsim 1$\,TeV.}
\end{figure}

\begin{figure}
\vspace{-1in}
\centerline{\epsfxsize=0.8\textwidth\epsfbox{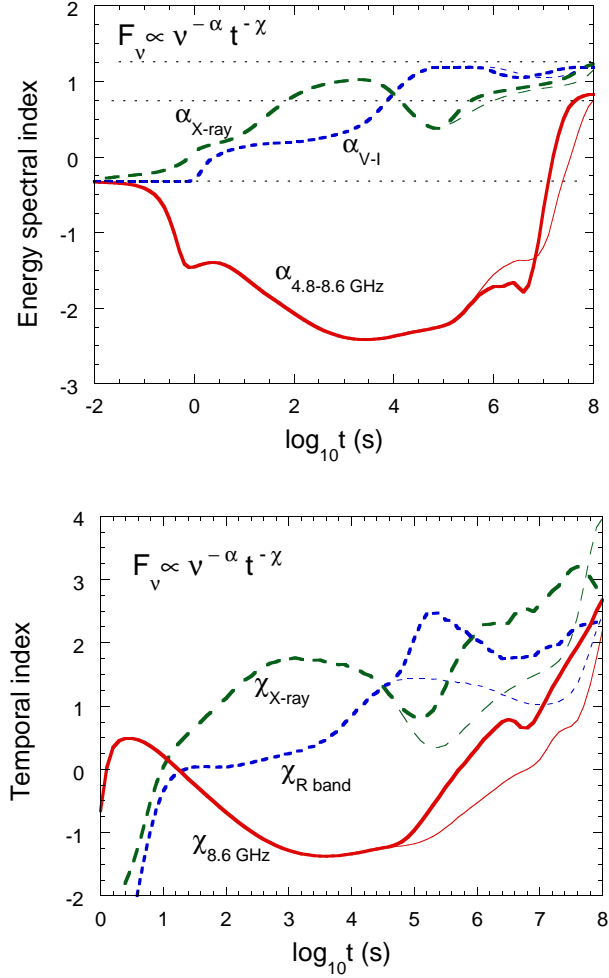}}
\vspace{0.5in}
\caption[]{Energy spectral indices and temporal indices for the 
uncollimated (light curves; from Fig.~\ref{standard_case}) and beamed
(heavy curves; from Fig.~\ref{lcs_std}) cases with $\psi = 10^\circ$
and $\theta = 0^\circ$. (a) Energy spectral indices $\alpha$
calculated between 4.8 and 8.6 GHz, the I and V bands, and 3 and 100
keV. (b) Temporal indices $\chi$ calculated at 8.6 GHz, the R band,
and 3 keV.}
\end{figure}

\begin{figure}
\vspace{-1in}
\centerline{\epsfxsize=0.8\textwidth\epsfbox{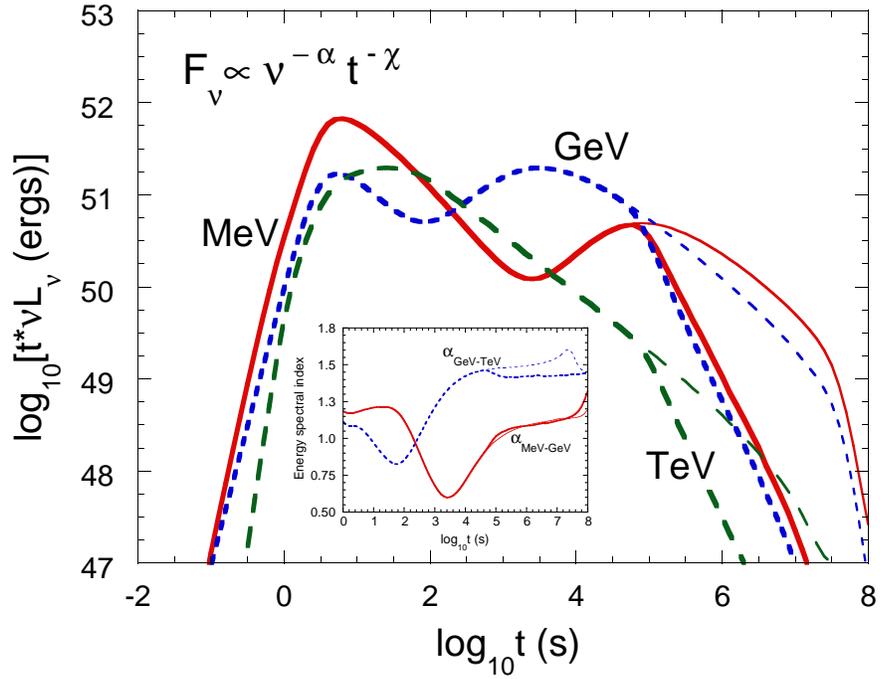}}
\vspace{0.5in}
\figcaption{Product of $\nu L_\nu$ flux and observing time $t$ for the 
MeV, GeV, and TeV light curves using the $\theta = 0^\circ$ case in
Fig.~\ref{lcs_std}b.  Inset shows temporal variation of the broadband
MeV--GeV and GeV--TeV energy spectral indices.}
\end{figure}

\begin{figure}
\centerline{\epsfxsize=0.5\textwidth\epsfbox{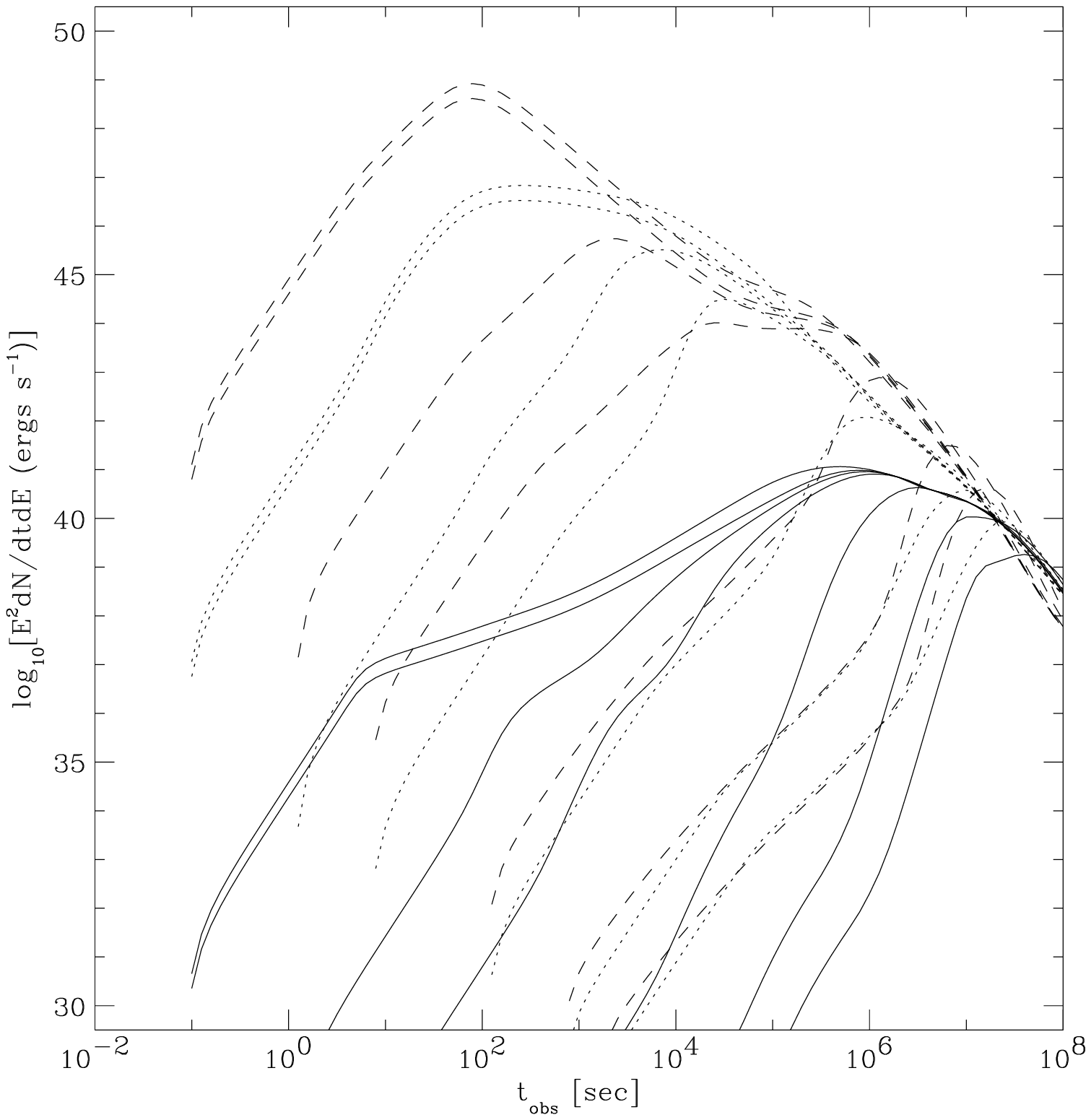}
            \epsfxsize=0.5\textwidth\epsfbox{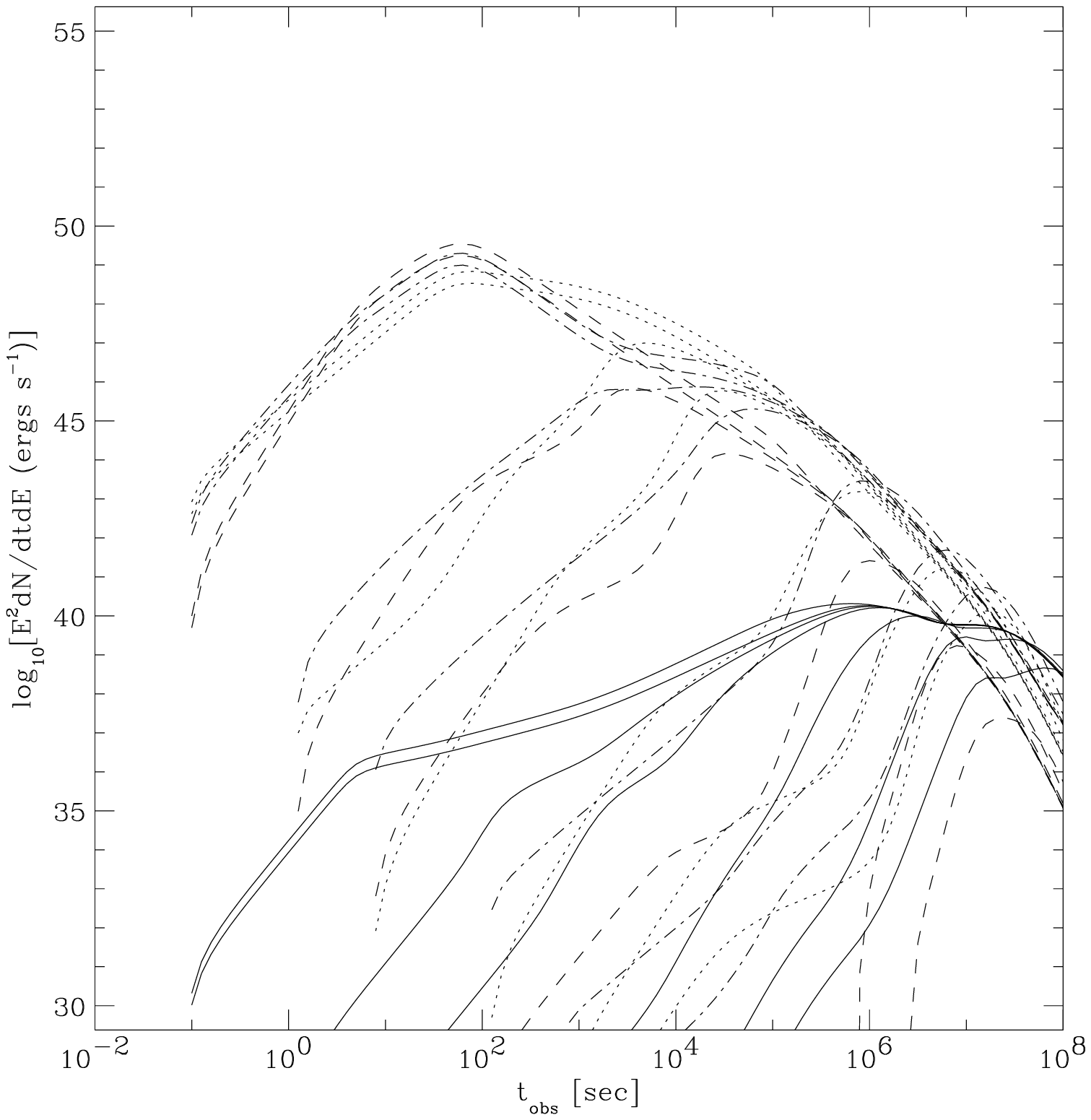}}
\caption{Same as Fig.~\ref{lcs_std}, but for an initial blast wave 
Lorentz factor $\Gamma_0 = 100$.}
\end{figure}

\begin{figure}
\centerline{\epsfxsize=0.5\textwidth\epsfbox{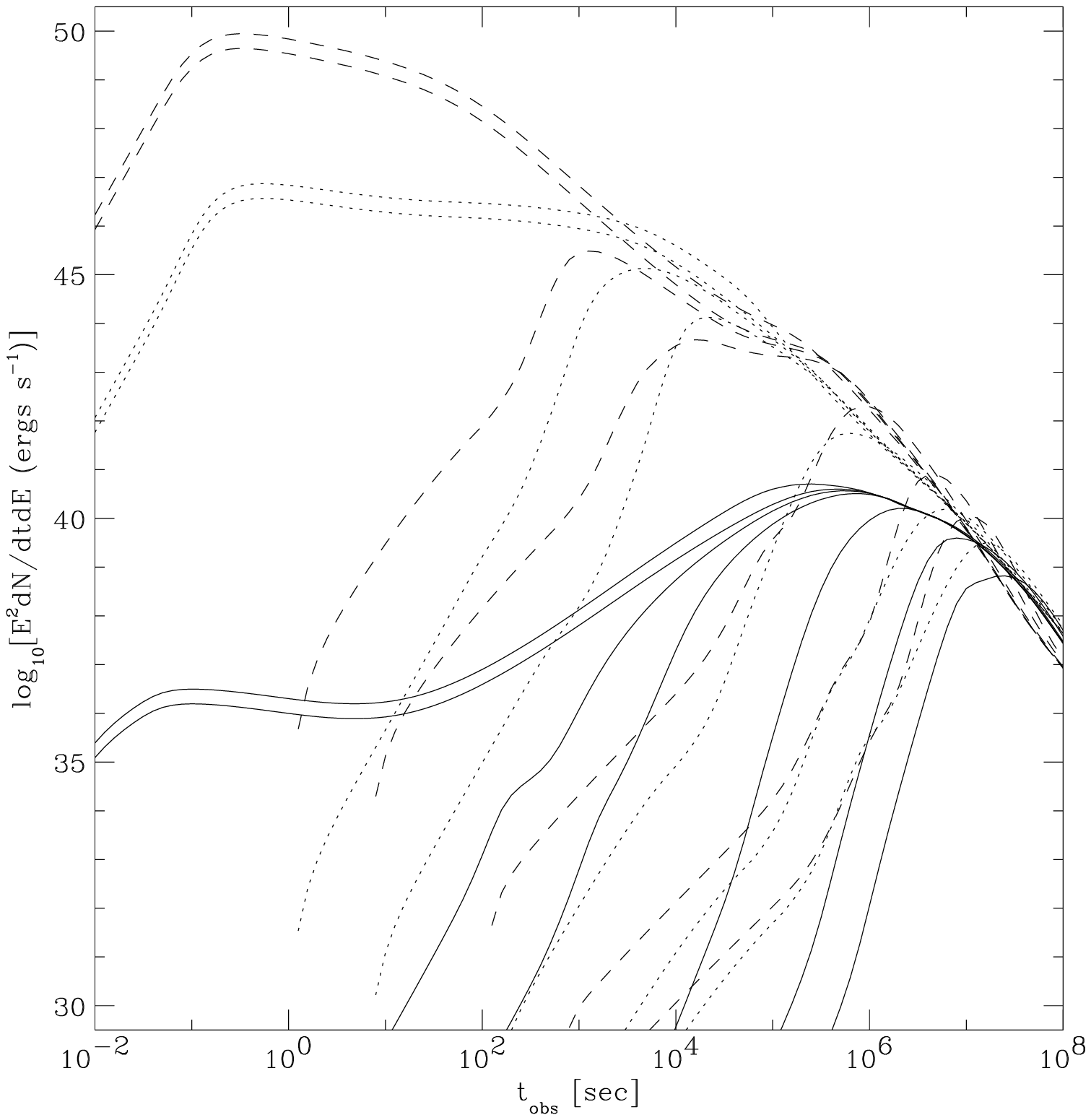}
            \epsfxsize=0.5\textwidth\epsfbox{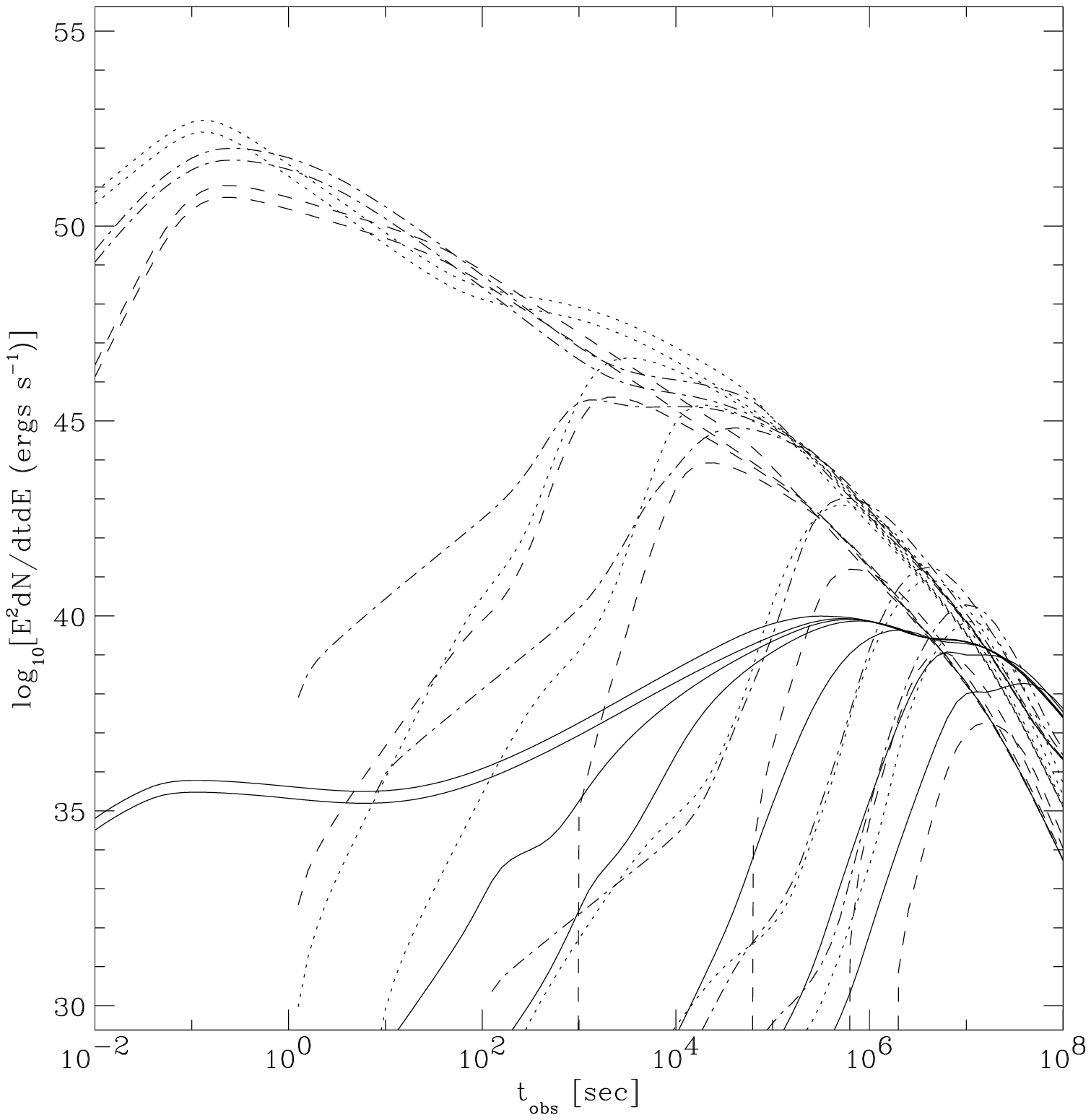}}
\caption{Same as Fig.~\ref{lcs_std}, but for an initial blast wave 
Lorentz factor $\Gamma_0 = 1000$.}
\end{figure}

\begin{figure}
\vspace{-1in}
\centerline{\epsfxsize=0.8\textwidth\epsfbox{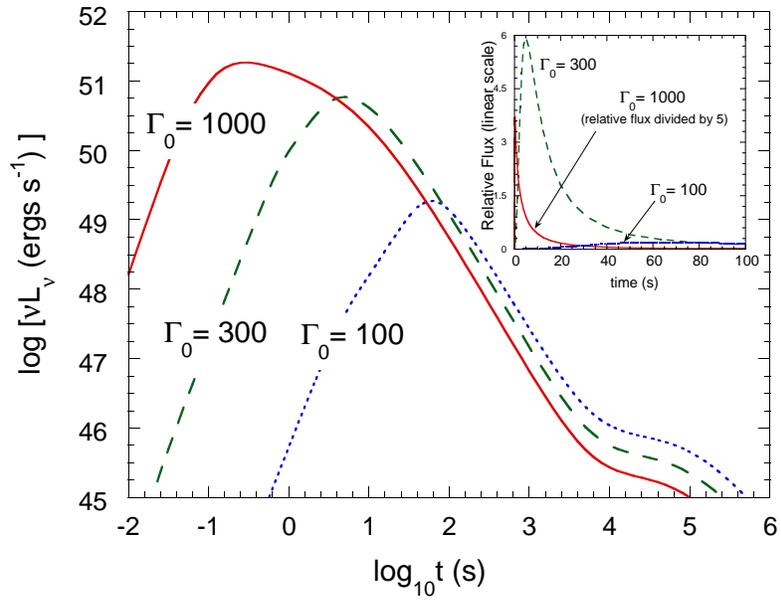}}
\vspace{0.5in}
\caption[]{Prompt and early afterglow 
100 keV light curves for blast wave Lorentz factors 
$\Gamma_0 = 1000$ (solid curve), 300 (dashed), and 100 (dotted).
Inset shows light curves on a linear scale.}
\end{figure}

\end{document}